\documentclass[groupedaddress,showpacs,showkeys,amssymb,eqsecnum,aps]{revtex4}
\usepackage[pdftex]{graphicx}
\usepackage{amsmath} 
\usepackage{epsf}                                                                                           
\usepackage{color}                   
\usepackage{verbatim}                                                                         

\newcommand{\be}{\begin{equation}}

\newcommand{\ee}{\end{equation}}

\begin{document}                                                                                              

\title{Renormalization of six-dimensional Yang-Mills theory in a background gauge field}    

\author{F. T. Brandt and J. Frenkel}
\email{fbrandt@usp.br, jfrenkel@if.usp.br}
\affiliation{Instituto de F\'{\i}sica, Universidade de S\~ao Paulo, S\~ao Paulo, SP 05508-090, Brazil}
\author{D. G. C. McKeon}
\email{dgmckeo2@uwo.ca}
\affiliation{
Department of Applied Mathematics, The University of Western Ontario, London, ON N6A 5B7, Canada}
\affiliation{Department of Mathematics and Computer Science, Algoma University,
Sault St.Marie, ON P6A 2G4, Canada}
                                                                                                              
\date{\today}

\begin{abstract}
Using the background field method, we study in a general covariant gauge the renormalization
of the 6-dimensional Yang-Mills theory. This requires background gauge invariant counterterms, some 
of which do not vanish on shell. Such counterterms occur, even off-shell, with gauge-independent 
coefficients. The analysis is done at one loop order and the extension to higher orders is discussed
by means of the BRST identities. We examine the behaviour of the beta function, which implies that this theory
is not asymptotically free.
\end{abstract}                                                                                                


\maketitle

\section{Introduction}

The background field formulation is a procedure which enables the calculation of quantum corrections,
while preserving the gauge invariance of the background field. This is an useful
method which has been much employed in non-abelian field theories,  like the Yang-Mills (YM) \cite{Dewitt:1967ub,KlubergStern:1974xv,Abbott:1980hw,McKeon:1994ds,Grassi:1995wr,weinberg:book95,Barvinsky:2017zlx,Batalin:2018enf,Frenkel:2018xup}
or gravity \cite{tHooft:1973bhk,Goroff:1986th,vandeVen:1992gw}  
gauge theories. In particular, it has been shown that on mass shell,  pure gravity
is renormalizable to one-loop order, despite the fact it contains a dimensional coupling which 
would make the theory non-renormalizable by power counting. This calculation has been done
in particular gauges, by using a 'topological invariant' which relates the scalars constructed from four derivatives of the gravitational field. It could be interesting
to extend this calculation to a general gauge,  but this would be very involved in the context of quantum gravity. 

The purpose of this work is to perform such an analysis in a simpler gauge model. To this end, 
we consider the conventional Yang-Mills (YM) theory in six dimensions, where the coupling constant has dimension of length, like in gravity.
The present analysis is done in a general covariant gauge which maintains the background gauge symmetry.            
The one-loop counter-terms have been earlier given in the Feynman gauge by van de Ven \cite{vandeVen:1984zk}.
Other aspects of the six-dimensional YM theory have been preciously studied from several points of view
\cite{vandeVen:1984zk,VanNieuwenhuizen:1977ca,Witten:1997kz,Saemann:2012rr,brs74,Gracey:2015xmw}.

Similarly to gravity, there exist Bianchi invariants connecting various counterterms involving 
four and higher derivatives of the YM field. But unlike the case of gravity, we find to one-loop order  
a counterterm
which does not vanish on mass shell and appears with a gauge-independent coefficient.
This means that the six-dimensional YM theory is not renormalizable in the power-counting sense. However,  we show
that it is renormalizable in the sense that there are gauge-invariant counterterms available to cancel all the ultraviolet divergences.
We also find that this YM theory is not asymptotically free, which is consistent with general arguments concerning non-renormalizable 
(by power counting) gauge theories
\cite{weinberg:book95,
  tHooft:1973mm}.

In section 2 we outline the background field method and give the basic Lagrangian which contains all interactions allowed,
to one-loop order, by the background gauge symmetry.  We study the renormalizability of the theory,  which requires the inclusion of two independent
counterterms involving four derivatives of the background field. The renormalization to higher orders is examined in section 4,  by means of a generalization
of the BRST identities. 

This symmetry, together with the background gauge invariance, is sufficient to ensure the 
renormalizability of the theory to all orders, in the more general sense. In section 4 we study the beta function ,  which is relevant for the asymptotic behavior,
and discuss its dependence on the definition of the running coupling.   
We analyse the gauge-independence of the 
coefficients of non-vanishing (on-shell) counter terms in section 5, where we give a summary of the results.
An outline of one-loop calculations in a general covariant gauge is presented in Appendix A
while other technical details are provided in the subsequent Appendices. 

\section{One-loop renormalization}

In the YM theory, the background field method is based on the gauge invariant Lagrangian 
\be\label{eq1}
{\cal L}_{YM}(A) = -\frac 1 4 \left(\partial_\mu A^{a}_\nu  - \partial_\nu A^{a}_\mu  + g f^{abc} A^{ b}_\mu A^{ c}_\nu \right)^2
\equiv -\frac 1 4 \left(F_{\mu\nu}^a(A)\right)^2
\ee
where $A_\mu^a$ is split into a background field $B_\mu^a$ and a quantum field $Q_\mu^a$, so $A_\mu^a   = B_\mu^a  + Q_\mu^a$.
The gauge-fixing Lagrangian is made to depend upon $B_\mu^a$ as
\be\label{eq2}
         {\cal L}_{GF} =  - \frac{1}{2\xi}\left(D_\mu(B) \cdot  Q^{\mu}\right)^2
\ee
where $\xi$ is a gauge-fixing parameter and $D_\mu(B)$ is the covariant derivative
\be\label{eq3}
D_\mu(B) =  \partial_\mu + g B_\mu \wedge .
\ee
Here we suppressed color indices and used the notation $B_\mu\cdot Q_\nu = B^a_\mu Q_\nu^a$; $(B_\mu\wedge Q_\nu)^a = f^{abc} B_\mu^b Q_\nu^c$
This gauge-fixing term leads to the following ghost Lagrangian
\be\label{eq4}
{\cal L}_{ghost} = -[\bar c D_\mu(B)]\cdot [D^\mu(B+Q) c].
\ee
Thus, the complete tree Lagrangian is given by
\be\label{eq5}
{\cal L}_{(0)} = {\cal L}_{YM}(B+Q) - \frac{1}{2\xi}\left[D_\mu(B)  Q^{\mu}\right]^2
-\left[\bar c D_\mu(B)\right]\cdot \left[D^\mu(B+Q)    c\right]. 
\ee
This Lagrangian has a background gauge symmetry under
\be\label{eq6}
\delta B_\mu = D_\mu(B) \,\omega ;\;\;\delta Q_\mu = g Q_\mu \wedge \omega;\;\;
\delta (c,\bar c)=g \, (c,\bar c) \wedge \omega.
\ee
where $\omega$ is an arbitrary infinitesimal parameter. 

We renormalize the theory to one loop order, by requiring the ultraviolet divergences of the background field amplitudes to
be cancelled by appropriate counter-terms. On dimensional and background gauge invariance grounds, one finds that the corresponding counter-term Lagrangian may have the structure
\be\label{eq7}
         {\cal L}_{(1)}(B) = 
           c_1 g^2\left(D_\sigma{ F}^{\sigma\mu}\right)^2 
+ c_2 g^3 { F}_{\mu}^{\;\;\sigma}\cdot( { F}_{\sigma\rho} \wedge { F}^{\rho\mu}) 
+ c_3 g^2 { F}^{\mu\nu}\cdot(D^2{ F}_{\mu\nu}) 
\ee
where 
$c_1$, $c_2$ and $c_3$ are dimensionless coefficients.
But, as shown in Appendix C, one finds a Bianchi identity which relates these terms as
\be\label{eq8}
(D^\sigma{ F}_{\sigma\mu})^2 +g { F}_{\mu}^{\;\;\sigma} \cdot({ F}_{\sigma\rho} \wedge { F}^{\rho\mu}) 
+ \frac 1 2 { F}^{\mu\nu}\cdot (D^2{ F}_{\mu\nu}) = \partial^\mu ({ F}_{\mu\nu}\cdot D_\sigma{ F}^{\sigma\nu}) 
\ee
Being a pure derivative, it follows that only two structures in Eq. \eqref{eq8} may be independent.
For definiteness, we can take these to be the first two terms in Eq. \eqref{eq8}. 
Hence, to one-loop order, the counter-term action may be written in the form
\be\label{eq9}
\Gamma_{(1)}^{CT}[B]= \int d^6 x \left\{
c_{11} g^2 \left(D_\sigma{ F}^{\sigma\mu}\right)^2 +  
c_{12} g^3 { F}_{\mu}^{\;\;\sigma}\cdot( { F}_{\sigma\rho} \wedge { F}^{\rho\mu}) \right\}
\ee
Using dimensional regularization in $d = 6 -2\epsilon$ dimensions, we have evaluated in a general covariant gauge
(see Eqs. \eqref{a26} and \eqref{a30} in Appendix A) the divergent coefficients $c_{11}$ and $c_{12}$.
To one-loop order, we have obtained that, for $SU(N)$ YM theory
\be\label{eq10}
c_{11} = -\frac{N}{32\pi^3\epsilon}\left(\frac{107}{240}-\frac{\xi}{8} -\frac{\xi^2}{48}\right);\;\;\; c_{12} = -\frac{N}{32\pi^3\epsilon}\frac{1}{180}
\ee
In consequence of the background gauge invariance, this effective action should obey the simple Ward identity
\be\label{eq11}
D_\mu(B)\cdot \frac{\delta \Gamma_{(1)}[B]}{\delta B_\mu} =  0.
\ee
The corresponding identities for the two and three point background field amplitudes have
been explicitly verified to one-loop order. 

We note here that the term proportional to $c_{11}$ in Eq. \eqref{eq9}  vanishes when the $B$-field equation
of motion is used.  This term may be removed by a non-linear field transformation in   ${\cal L}_{YM}(B)$
\be\label{eq12}
B_\mu \rightarrow B_\mu - c_{11}  g^2 D^\sigma(B) F_{\sigma\mu}(B).
\ee
Such a field redefinition has no physical observable effects. Thus, there is no reason why the coefficient of this structure should be independent of the gauge parameter.
On the other hand,  the coefficient c12 of the structure which does not vanish “on-shell” is gauge-independent. This
result is necessary in order to ensure the gauge-independence of physical S-matrix elements, 
as discussed in Appendix B.

\section{Renormalization to higher orders}

In order to cancel the ultraviolet divergences of the background field amplitudes at two loops, 
we must add an appropriate set of independent gauge-invariant counter-terms.  As discussed
in Appendix C,  the Bianchi identities lead to a suitable set of the form
\begin{eqnarray}\label{eq13}
   {\cal L}^{CT}_{(2)}(B) &=& 
c_{21} g^4 \left[D_\mu D_\sigma{ F}^{\sigma\nu}\right]^2 +  c_{22} g^5 (D_\sigma{ F}^{\sigma\mu})
\wedge(D_\rho{ F}^{\rho\nu})\cdot { F}_{\mu\nu}
+  c_{23} g^5 (D^\mu D^\sigma{ F}_{\sigma\nu})\cdot( { F}^{\nu\rho} \wedge {\cal F}_{\rho\mu})
\ \nonumber \\
&& +  c_{24} g^6 \left[\left({ F}_{\mu\nu}\right)^2\right]^2
+ c_{25} g^6 \left[{ F}_{\mu\nu} \cdot { F}_{\rho\sigma}\right]^2
+ c_{26} g^6 \left[{ F}_{\mu\nu} \cdot { F}_{\nu\rho}\right]^2
+ c_{27} g^6 (F_{\mu\sigma} \wedge F^{\sigma\rho})\cdot(F_{\rho\nu} \wedge F^{\nu\mu})
\end{eqnarray}
where $c_{2i}$ are real dimensionless coefficients.
We remark that, on shell, we need four new counter-terms, compared with the pure gravity at two loops which requires just one new counter-term.

We must show that the theory can be renormalized
in the sense that there is a counter-term available
to cancel every ultraviolet divergence,
in a way which preserves the background gauge invariance.
To this end, one must also use the Becchi-Rouet-Stora-Tyutin (BRST) symmetry.
We note that when the background method is used to two-loop order or higher, the subgraphs
become functionals of $B_\mu$ as well as of $Q_\mu$, $c$ and $\bar c$, leading to an effective action $\Gamma[B,Q, c, \bar c]$ which has a background gauge invariance under Eq. \eqref{eq6}.

We remark that ${\cal L}_{(0)}$  in  \eqref{eq5}   with the gauge-fixing term left out, is also invariant under the BRST transformations
\be\label{eq15}
\Delta B = 0;\;\Delta Q_\mu = D_\mu(B+Q) c \tau;\; \Delta c = -\frac 1 2 g c\wedge c \tau;\; \Delta \bar c = 0
\ee
where $\tau$ is an infinitesimal anti-commuting constant. Thus, we have that 
\be\label{eq16}
\int d^6 x
\left[
\frac{\delta \Gamma_{(0)}^\prime}{ \delta Q_\mu}\cdot \Delta Q_\mu+
\frac{\delta \Gamma_{(0)}^\prime}{ \delta c}\cdot \Delta c \right]=0
\ee
where a prime denotes the fact that the gauge-fixing term has been left out. 

One may write this invariance in a more useful form, by introducing the Zinn-Justin source-terms $U_\mu$, $V$ \cite{Zinn-Justin:1974mc}
which are coupled respectively
to the BSRT variations $\Delta Q_\mu$ and   $\Delta c$     as      
\be\label{eq17}
{\cal L}_{ZJ} = U^\mu \Delta Q_\mu + V \Delta c = U^\mu D_\mu(B+Q) c - \frac 1 2 V\cdot (c\wedge c).
\ee
One may verify that this Lagrangian is also invariant under the BRST transformations \eqref{eq15}. 

Adding ${\cal L}_{ZJ}$  to ${\cal L}_{(0)}$  in \eqref{eq5},  with the gauge fixing term omitted, and denoting
by $\tilde{\cal L}_{(0)}$  the total tree Lagrangian thus obtained, one gets from \eqref{eq16} the lowest order Zinn-Justin equations
\be\label{eq18}
\int d^6 x
\left[\frac{\delta \tilde\Gamma_{(0)}}{ \delta Q_\mu}\cdot \frac{\delta \tilde\Gamma_{(0)}}{\delta U^\mu}+
\frac{\delta \tilde\Gamma_{(0)}}{ \delta c}\cdot \frac{\delta \tilde\Gamma_{(0)}}{ \delta V}
\right]=0
\ee
which may be extended to all orders. These identities,  together with the background gauge invariance, are sufficient to fix the renormalization
of the YM theories which are renormalizable by power-counting. 

However, for gauge theories which are non-renormalizable by power-counting , the proof of renormalizability in the more general sense,
requires a generalization of Zinn-Justin method,  known as the Batalin-Vilkovisky formalism.
In this formalism one includes, apart from the linear source-terms in \eqref{eq17}
also non-linear functionals of the sources, which have ghost number zero
\cite{Batalin:1981jr,Batalin:1984ss}.
The inclusion of such supplementary terms preserves the master equation
\be\label{eq19}
\int d^6 x
\left[\frac{\delta \tilde\Gamma}{ \delta Q_\mu}\cdot \frac{\delta \tilde\Gamma}{\delta U^\mu}+
\frac{\delta \tilde\Gamma}{ \delta c}\cdot \frac{\delta \tilde\Gamma}{ \delta V}
\right]=0.
\ee
This equation reflects the invariance of the action under a generalized BRST transformation. 
The relevance of the Batalin-Vilkovisky method is that all the consistency conditions for this
generalized symmetry are incorporated in the master equation. 


Defining
\be\label{3.7a}
\tilde\Gamma^\prime\star \tilde\Gamma^{\prime\prime} =
\int d^6 x\left[
\frac{\delta \tilde\Gamma^\prime}{\delta Q_\mu}\cdot\frac{\delta \tilde\Gamma^{\prime\prime}}{\delta U^\mu}+
\frac{\delta \tilde\Gamma^\prime}{\delta c}\cdot\frac{\delta \tilde\Gamma^{\prime\prime}}{\delta V}
\right]
\ee
we can write \eqref{eq19} in the concise form
\be\label{3.7b}
\tilde\Gamma\star\tilde\Gamma = 0.
\ee
Let us now perform a loop expansion on the generating functional $\tilde\Gamma = \sum_{n} \tilde\Gamma_{(n)}$.
Substituting this in \eqref{3.7b}, gives to second order
\be\label{3.7c}
\tilde\Gamma_{(0)}\star \tilde\Gamma_{(2)}+\tilde\Gamma_{(2)}\star \tilde\Gamma_{(0)}+\tilde\Gamma_{(1)}\star \tilde\Gamma_{(1)}=0 
\ee
where $\tilde\Gamma_{(1)}$ was made finite by the addition of the
counterterm \eqref{eq9}. Since $\tilde\Gamma_{(0)}$ is finite, the
divergent part of $\tilde\Gamma_{(2)}$ satisfies
 \be\label{3.7d}
\tilde\Gamma_{(0)}\star \tilde\Gamma_{(2)}^{div}+\tilde\Gamma_{(2)}^{div}\star \tilde\Gamma_{(0)}=0.
\ee
Using \eqref{3.7a}, this leads to the relation
\begin{eqnarray}\label{3.7e}
\int d^6 x  \left[\left(\frac{\delta\tilde\Gamma_{(0)}}{\delta Q_\mu}\cdot \frac{\delta}{\delta U^\mu}
                              +\frac{\delta\tilde\Gamma_{(0)}}{\delta c}\cdot \frac{\delta}{\delta V}\right)
+\left(\frac{\delta\tilde\Gamma_{(0)}}{\delta U^\mu}\cdot \frac{\delta}{\delta Q_\mu}
                              +\frac{\delta\tilde\Gamma_{(0)}}{\delta V}\cdot \frac{\delta}{\delta c}\right)\right]
\tilde\Gamma_{(2)}^{div}  \equiv {\cal G} \tilde\Gamma_{(2)}^{div} = 0,
\end{eqnarray}
Here ${\cal G}$ has the important property that it is idempotent:
${\cal G}^2=0$. This implies that a solution of \eqref{3.7e} may be
written as 
\be\label{3.7f}
\tilde\Gamma^{div}_{(2)} = {\cal G} F(B,Q,c,\bar c;U,V) + G[B,Q],
\ee
where $F$ is some function and $G[B,Q]$ is a generic
gauge-invariant functional.

But, unlike the case in renormalizable theories, this solution is not
sufficient to generate all the ultraviolet divergences at 2 loops.
In order to account for all infinities, one must also include a set of
renormalized fields and sources, defined in terms of the original
quantities by a general canonical transformation
\cite{weinberg:book95}. This procedure yields, in a way that preserves
the background gauge invariance, all the counter-terms
$\tilde\Gamma_{(2)}^{CT} = -\tilde\Gamma_{(2)}^{div}$ which are needed
to cancel every divergence \cite{Barvinsky:2017zlx}. Integrating out
the fields $Q$, $\bar c$, $c$ and dropping the auxiliary sources $U$,
$V$, leads at two loop order, to the general counter-term action
$\tilde\Gamma_{(2)}^{CT}[B]$ which has a background gauge symmetry
\be\label{3.7g}
\tilde\Gamma_{(2)}^{CT}[B] = \int d^6 x {\cal L}_{(2)}^{CT}(B),
\ee
where ${\cal L}_{2}^{CT}$ is given in \eqref{eq13}. 
The proof that this operation holds to all orders may be made 
recursively by introducing, order by order, appropriate sets of 
renormalized fields and sources, as has been argued in \cite{Gomis:1996jp}.

Using this  result, we still need to examine a subtlety which arises in the background 
gauge formalism. In this method, it is necessary to omit certain one-particle reducible graphs involving vertices which are linear in $Q$
\cite{Abbott:1980hw,weinberg:book95,Frenkel:2018xup}.
Although the omission of such terms preserves the background gauge invariance,  this violates the BRST symmetry.
Thus, the correct effective action for the background field, is not BRST invariant.  However,
the BRST invariance of the original action $\Gamma[B,Q,c,\bar c]$,  can be used in an indirect way to control the renormalization of      
$\bar\Gamma[B,Q,c,\bar c]$. To this end, we may use the generalized BRST approach to renormalize $\Gamma[B,Q,c,\bar c]$
and then deduce the renormalization of the background effective action $\Gamma[B,Q,c,\bar c]$
by the operation \cite{Frenkel:2018xup}
\be\label{eq20}
\bar\Gamma_R = \Omega \Gamma_R \equiv \Gamma_R - \int d^6 x \, 
Q_\mu\cdot \left[\frac{\delta\Gamma_R}{\delta Q_\mu}\right]_{Q=c=0} 
\ee
where the terms which are linear in $Q_\mu$, but independent of $c$,  have been subtracted.

\section{The beta function}
The renormalized coupling constant $g(\mu)$  is usually  defined in terms of the value of some
physical process, which is evaluated at a characteristic energy of magnitude $\mu$.  For example, 
one may consider the gluon-gluon scattering amplitude: $p_1 + p_2 \rightarrow p_3 + p_4$ , where $\mu$ may be
identified with the total energy of the system in the center of mass frame. 

The one-loop divergent contributions to this physical amplitude, involving structures like those
in the first counterterm in \eqref{eq9}, cancel among themselves (see Appendix B). Thus, such a counterterm
is superfluous in this case, as one may expect from the discussion at the end of section 2. On
the other hand, the contributions arising from the second counterterm in \eqref{eq9}, with occurs with
the coefficient $c_{12}$, are necessary to cancel out the remaining  divergences in the one-loop amplitude. 
In the six-dimensional theory the form of the divergent part of the one-loop
amplitude is different from that of the tree-level amplitude.
However, evaluating these amplitudes with transverse polarization vectors, it turns out that at the point
\be\label{eq21}
p_1^2 = p_2^2 = p_3^2 =p_4^2 = 0; \;\;\;  (p_1 +p_2)^2 = -2 (p_1 -p_3)^2 = -2 (p_1 -p_4)^2 = \mu^2 
\ee
the divergent part of the one-loop amplitude becomes proportional to the bare tree amplitude
evaluated from the first term in Eq. \eqref{eq9}, with a bare coupling $g_0$. This fact enables
to connect the bare coupling 
to the one-loop divergences, by requiring the sum of these two amplitudes to be finite.
The value of the total physical amplitude
at the point \eqref{eq21} may then be used to define the renormalized coupling constant $g(\mu)$.

This procedure allows us to relate the bare coupling $g_0$
to the renormalized coupling $g(\mu)$, through a factor which involves the divergent coefficient $c_{12}$. To this end, we note that
the ultraviolet divergences in the loops arise as powers of poles $1/\epsilon$.  
To cancel these poles,  the bare coupling $g_0$ must itself have such poles. 
Thus, the coupling $g_0$  may be expressed in terms of a series of powers of $1/\epsilon$  as
\be\label{eq22}
g_0 = g(\mu)\left[1+c_1[g(\mu)]/\epsilon + \dots\right]
\ee
where, to one loop order, the dimensionless factor $c_1/\epsilon$ is proportional  to  $c_{12}\mu^2 g^2(\mu)$.                      

We note that in $d = 6 - 2\epsilon$ dimensions, the mass dimension of
$g (\mu,\epsilon)$  is $\epsilon-1$.  It is convenient 
to rescale this dimensionful coupling as
\be\label{eq23}
g(\mu,\epsilon) \equiv \mu^{\epsilon-1} \tilde g(\mu,\epsilon)
\ee
where $\tilde g(\mu,\epsilon)$, is a dimensionless renormalized coupling constant.  

Thus, we may rewrite \eqref{eq22} in the form
\be\label{eq24}
g_0 =
(\mu)^{\epsilon-1}\tilde g(\mu,\epsilon)\left[1+\frac{c_1(\tilde g)}{\epsilon}+\frac{c_2(\tilde g)}{\epsilon^2} + \dots 
\right]  \equiv (\mu)^{\epsilon-1} \tilde g(\mu,\epsilon) \tilde Z[\tilde g(\mu,\epsilon)]
\ee
where,  to lowest order, $c_1(\tilde g)/\epsilon \approx c_{12} \tilde g^2$.

An important property of the parameter $\tilde Z$ is that it is a gauge-independent
quantity,  when using the minimal subtraction renormalization scheme. To see this, we note
that since the bare coupling constant $g_0$ and $\mu$ are independent of the gauge parameter, 
we have from \eqref{eq24}
\be\label{eq25}
\frac{d}{d\xi} (\tilde g \tilde Z) = 0.
\ee
Inserting $\tilde Z$ defined by \eqref{eq24} into \eqref{eq25} and grouping  the terms in powers  of $1/\epsilon$,  we get
\be\label{eq26}
 \frac{d\tilde g}{d\xi} + \frac{1}{\epsilon} \left(c_1\frac{d\tilde g}{d\xi} + \frac{dc_1}{d\xi}\tilde g\right)+\dots = 0
\ee
Since the terms of different powers of $1/\epsilon$ in \eqref{eq26} are independent (because the 
coefficients of these powers are finite as    $\epsilon\rightarrow 0$) each term should vanish separately, so that 
\be\label{eq27}
\frac{d\tilde g}{d\xi}=0;\;\;\; \frac{d c_1}{d\xi}=0;\;\;\; \dots 
\ee
Thus,  we find that $\tilde g$  as well as   $\tilde Z$ must be gauge-independent quantities. In particular, 
this explains the fact that coefficient $c_{12}$ in \eqref{eq10}  is independent of the gauge-parameter $\xi$. 

To calculate the beta-function,  one differentiates \eqref{eq24}  with respect to  $\mu$  and use the fact
that the bare coupling is independent of $\mu$. One then gets the relation 
\be\label{eq28}
\tilde Z^{-1} \mu^{2-\epsilon}\frac{dg_0}{d\mu} = 0 = (\epsilon-1)\tilde g(\mu,\epsilon)
+\beta[\tilde g(\mu,\epsilon)]\left\{1+\tilde g(\mu,\epsilon) \frac{d\log(\tilde Z)}{d\tilde g(\mu,\epsilon)}\right\}
\ee
where we have used the chain rule and defined
\be\label{eq29}
\beta[\tilde g(\mu,\epsilon)] \equiv \mu \frac{d}{d\mu} \tilde g(\mu,\epsilon)
\ee

Since the beta-function should be finite, the coefficients of the various powers of $1/\epsilon$ in  
$\tilde Z$ must be related in such a way  so as to ensure the cancellation of pole terms  in \eqref{eq28} .  
To get the beta function, one expands all quantities in powers of  $\epsilon$    
and equate to zero the terms of zeroth and first order in $\epsilon$. 
Then, taking the limit  $\epsilon\rightarrow 0$,  one finds the simple relation 
\be\label{eq30}
\beta[\tilde g(\mu)] = \tilde g(\mu) + \left[\tilde g(\mu) \frac{d}{d \tilde g(\mu) }-1\right]  \tilde g(\mu)  c_1[\tilde g(\mu) ],
\ee
which involves only the coefficient $c_1$ of the simple pole in \eqref{eq24}.

We show in Appendix D that the form of the first term  in the power series for the beta function 
is independent  of the definition of the running coupling. But this is generally not the case for the higher order terms.   
However,  this arbitrariness is not important for small couplings,  since then it is just the first term which 
determines the leading behaviour of the beta function. 
It then follows from \eqref{eq29}  and \eqref{eq30}  that for small couplings,  $\tilde g(\mu)$ grows linearly 
with $\mu$. 
One may note, upon inverting \eqref{eq23}, that $\tilde g$ is a function of $\mu$ as well as of the
dimensional coupling g.
Thus, the original renormalized coupling $g(\mu)$ in \eqref{eq23} becomes (up to small corrections), 
independent  of the scale $\mu$. This implies that the six-dimensional Yang-Mills theory is not asymptotically free. 

\section{Conclusion}

We have examined the renormalization of the six-dimensional Yang-Mills theory, which has a coupling with length dimension,
as a model for the gauge theory of gravity. The YM theory was
studied in a general covariant gauge which preserves the background field invariance.To one-
loop order we find, similarly to pure gravity which has been studied in particular gauges, that
there appear counterterms which vanish on shell, with gauge-dependent coefficients. But such
terms are unphysical, since they can be turned away by a field redefinition (Eq. \eqref{eq12}).
On the other hand, unlike pure gravity where all the counterterms vanish on shell, we get a non-vanishing 
counterterm on shell, which occurs with a gauge-independent coefficient even off-shell
(Eqs. \eqref{eq9} and \eqref{eq10}).
This result was verified in a general background gauge,
to one loop-order. At two-loops, the renormalizability of the theory
requires four more counterterms which do not vanish
on-shell (Eq. \eqref{eq13}), as compared with pure gravity where only
one such counterterm is needed.

We note here that the gauge-independence of the coefficients of the counterterms which 
do not vanish on shell is an outcome of the background field method. To understand this, let us
us compare the present theory with the 4-dimensional YM theory, which is renormalizable by
power counting. The usual theory is multiplicatively renormalizable because the divergent part of Green’s functions
has the same form as the tree functions. In this case the bare coupling and 
the bare field may generally be related to the renormalized quantities by simple rescalings
\be
g_0 = Z_g g;\;\;\; B_0 = Z_B^{1/2} B.
\ee
In order to preserve the background gauge invariance,
the renormalization constants $Z_g$ and $Z_B$ must be connected, which leads to the renormalized Lagrangian
\be
{\cal L}_{YM}^{4\,R}(B) = -\frac 1 4 Z_B (F_{\mu\nu}(B)]^2 = -\frac 1 4 Z_g^{-2} (F_{\mu\nu}(B)]^2 
\ee
where $Z_g$ is a gauge-independent quantity (see Eqs. \eqref{a26a} and \eqref{a29})

On the other hand, the 6-dimensional YM theory is not multiplicatively renormalizable, since
the form of the divergent terms at $(n+1)$ loops is, in general, quite different from the one which
arise at $n$-loops. Instead, the theory is renormalizable in the more general sense that there 
are counterterms available to cancel all ultraviolet divergences \cite{Gomis:1996jp}.
For example, considering only the independent non-vanishing (on shell) counterterms and using Eqs. \eqref{eq9} and \eqref{eq13},
we may write the counter-term (up to two loops) Lagrangian in the form
\begin{eqnarray}\label{eq53}
         {\cal L}_{YM}^{6\, CT}(B) &=& 
c_{12} g^3 F_\mu^\sigma \cdot (F_{\sigma\rho} \wedge F^{\rho\mu}) + c_{24} g^6 [(F_{\mu\nu})^2]^2
\nonumber \\ &&
+c_{25} g^6 [F_{\mu\nu} \cdot F_{\rho\sigma}]^2 + c_{26} g^6 [F_{\mu\nu} \cdot F^{\nu}_{\;\rho}]^2
+ c_{27} g^6 (F_{\mu\sigma} \wedge F^{\sigma\rho})\cdot(F_{\rho\nu} \wedge F^{\nu\mu}).
\end{eqnarray}
This involves only the field strength $F_{\mu\nu}(B)$ of the background field and is manifestly
invariant under background gauge transformations.

As we have pointed out,
the counterterms which vanish on shell are unphysical and decouple in observable processes.
In physical amplitudes, only the counterterms which do not vanish on-shell are necessary to
cancel the loop-divergences (see Appendix B).
%
%
Using the Ward identities, one can show that the only possible gauge-dependence in these amplitudes may arise just from the coefficients
of these counterterms. But physical processes are generally gauge-invariant, a feature which must hold in particular
in the background field approach \cite{Abbott:1983zw}. Thus, in order to ensure this property, it follows that
the coefficients 
of the counterterms which
do not vanish on-shell  
should be gauge-independent quantities. This explains the explicit result obtained at one-loop order in \eqref{a27} and \eqref{a30}.
%
%
%

%

A further argument, based on the renormalization group method, for the
gauge independence of the coefficients of non-vanishing (on shell)
counterterms is given in section 4 (see Eq. \eqref{eq27}).
Here,  we have derived a  beta function which encodes the dependence of the running coupling $g(\mu)$ on the energy scale $\mu$
(Eq. \eqref{eq30}).
%
From this, we conclude that the six-dimensional YM theory is not asymptotically free, as expected for field theories
which are not renormalizable by power counting.

We presume that a similar behaviour also occurs in pure gravity at two
loops order, which requires a new counterterm that is cubic in the
curvature tensor
\cite{tHooft:1973bhk,Goroff:1986th,vandeVen:1992gw}. We expect 
the coefficient of this counterterm, which does not vanish on mass
shell, to be gauge independent. This may hold since such counterterms
are relevant for the renormalization of physical quantities.

\newpage

\appendix 

\section{Perturbative Calculations}
\subsection{Feynman Rules}
Using the tree Lagrangian given by Eq. \eqref{eq5} it is straightforward to obtain 
the momentum space Feynman rules which arises from $i S_{(0)} = i \int d^d x \,{\cal L}_{(0)}$.
The gluon 
and ghost propagators are given respectively by 
\begin{eqnarray}
\begin{array}{lcc}\label{a1} 
\input{QQ_AbbottFix.pspdftex}   
\end{array} \;\;\;\;\;\;  &=& \displaystyle{\frac{-i\delta^{ab}}{k^2+i0}\left[\eta_{\mu\nu}-(1-\xi)\frac{k_\mu k_\nu}{k^2+i0}\right]},
\\ 
\begin{array}{lcc}\label{a2} 
\input{cc_Abbott.pspdftex} 
\end{array} \;\;\;\;  &=& \displaystyle{
 \frac{i\delta^{ab}}{k^2+i 0}  } . 
\end{eqnarray} 
The interaction vertices which are relevant for the one-loop contribution  
to the two- and three-gluon background field functions (see figures \ref{figSE} and \ref{figThree}) are
\begin{eqnarray} 
\begin{array}{lcc}\label{a3}
\input{BQQ_Abbott.pspdftex} 
\end{array}  \;\;\; &=& 
\displaystyle{-g f^{abc} \left[\eta_{\nu\lambda}(r_\mu-q_\mu)+
                                         \eta_{\mu\lambda}(p_\nu-r_\nu-\frac{1}{\xi} q_\nu) 
                                        -\eta_{\mu\nu}(p_\lambda-q_\lambda-\frac{1}{\xi} r_\lambda) 
\right] },
\\
\begin{array}{lcc}\label{a4}
\input{BBQQ_Abbott.pspdftex} 
\end{array}  \;\;\; &=&  
\begin{array}{l}
\displaystyle{ -ig^2\bigg[
f^{abe} f^{cde} (\eta_{\mu\lambda} \eta_{\nu\rho} - \eta_{\mu\rho} \eta_{\nu\lambda} )+}\\
\;\;\;\;\;\;\;\;\;\;\displaystyle{
f^{ace} f^{bde} (\eta_{\mu\nu} \eta_{\lambda\rho} - \eta_{\mu\rho} \eta_{\nu\lambda} +\frac{1}{\xi}\eta_{\mu\lambda}\eta_{\nu\rho})+}\\
\;\;\;\;\;\;\;\;\;\;\displaystyle{
f^{bce} f^{ade} (\eta_{\mu\nu} \eta_{\lambda\rho} - \eta_{\nu\rho} \eta_{\mu\lambda} +\frac{1}{\xi}\eta_{\nu\lambda}\eta_{\mu\rho})
\bigg]},
\end{array}
\\
\begin{array}{lcc}\label{a5}
\input{Bcc_Abbott.pspdftex} 
\end{array}  \;\;\; &=& 
\displaystyle{-g f^{abc} (p_\mu+q_\mu) },
\\
\begin{array}{lcc}\label{a6}
\input{BBcc_Abbott.pspdftex} 
\end{array}  \;\;\; &=& 
\displaystyle{i g^2 \eta_{\mu\nu}( f^{ace} f^{bde} + f^{ade} f^{bce})}. 
\end{eqnarray}
where all momenta are oriented inwards and a blob indicates a background field.  
Vertices with all the external fields of the same type ($B$ or $Q$)
can be obtained from Eqs. \eqref{a3} and \eqref{a4} by setting $1/\xi=0$.

\subsection{Two- and Three-gluon functions at one-loop order}
\begin{figure}[t]
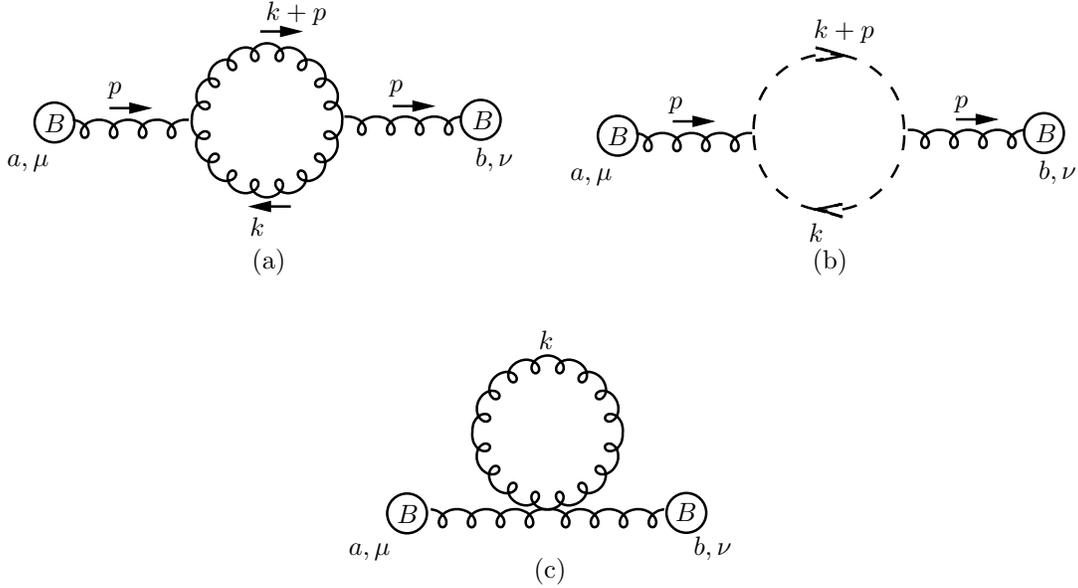

\begin{eqnarray}
\input{SE1_AbbottNew.pspdftex }  & \qquad \input{SE2_AbbottNew.pspdftex} 
 \nonumber 
\end{eqnarray}
\begin{eqnarray}
   \input{SE3_AbbottNew.pspdftex} 
\nonumber 
\end{eqnarray}
\caption{One-loop contributions to $\langle BB \rangle$.}
\label{figSE}
\end{figure}

The one-loop contributions to the two-point function $\langle BB \rangle$ are given
by the Feynman diagrams of Fig. \ref{figSE}.  Since the diagram in Fig. \ref{figSE} (c)  vanishes in dimensional
regularization, we only have to compute diagrams (a) and (b).
After the loop momentum integration,  the result can only depend (by covariance) on the two tensors
$\eta_{\mu\nu}$ and $p_\mu p_\nu$. A convenient tensor basis in terms of these tensors is 
\be
   {\cal T}_{\mu\nu}^1   = p_\mu p_\nu - k^2 \eta_{\mu\nu} \;\; \mbox{and} \;\; {\cal T}_{\mu\nu}^2 = p_\mu p_\nu
\ee
so that the diagrams in figure \ref{figSE}  can be written as $\Pi^{I\, ab}_{\mu\nu}(p)  = N g^2 \delta^{ab} \Pi^{I}_{\mu\nu}(p)$
(we are using $f^{amn} f^{bmn} = N \delta^{ab}$), where
\begin{equation}\label{eq2a}
\Pi^{I}_{\mu\nu}(p) =  \sum_{i=1}^{2}  {\cal T} ^i_{\mu\nu}(p) 
C^I_i(p) ; \;\;\; I=\mbox{a}\; \mbox{and}\; \mbox{b}.
\end{equation}
The coefficients $C^I_i$ can be obtained solving the following system of two algebraic equations
\be\label{ccc1}
\sum_{i=1}^2 {\cal T}^i_{\mu\nu}(p) {\cal T}^j{}^{\mu\nu}(p) C^I_i(p) =
\Pi^I_{\mu\nu}(p) {\cal T}^j{}^{\mu\nu}(p) \equiv J^I{}^j(p); \;\;
j=1,2 .
\ee

Using the Feynman rules of the previous section, it is straightforward to obtain the expressions for
each $\Pi^I_{\mu\nu}(p)$. Then, the integrals on the right hand side of \eqref{ccc1} have the following form
\be
J^I{}^j(p) = \int \frac{d^d k}{(2 \pi)^d}  s^I{}^j(k,q,p).
\ee
where $q=k+p$; $k$ is the loop momentum, $p$ is the external momentum and $s^I{}^j(k,q,p)$ are
scalar functions. Using the relations 
\begin{subequations}
\begin{eqnarray}
k\cdot p = (q^2 - k^2 - p^2)/2, \\
q\cdot p = (q^2 + p^2 - k^2)/2, \\
k\cdot q = (k^2 + q^2 - p^2)/2, 
\end{eqnarray}
\end{subequations}
the scalars  $s^I{}^j(k,q,p)$ can be reduced to combinations of powers of $k^2$ and $q^2$. As a result, the
integrals $J^I{}^j(p)$  can be expressed in terms of combinations of the following well known integrals 
\begin{equation}\label{scalInt}
I^{l m} \equiv 
\int \frac{d^d k}{(2 \pi)^d} \frac{1}{(k^2)^l (q^2)^m} = i \frac{(k^2)^{d/2-l-m}}{(4\pi)^{d/2}}
\frac{\Gamma(l+m-d/2)}{\Gamma(l) \Gamma(m)} \frac{\Gamma(d/2-l) \Gamma(d/2-m)}{\Gamma(d-l-m)},
\end{equation}
where  powers $l$ and $m$ greater than one may only arise from the terms proportional to $1-\xi$ in the gluon propagator 
(see Eq. \eqref{a1}). 
The only non-vanishing (ie non tadpole) integrals are 
%
\begin{subequations}\label{intregd}
\begin{eqnarray}
I^{11} & = & i \frac{(k^2)^{d/2-2}}{2^d\pi^{d/2}}
\frac{\Gamma \left(\frac{d}{2}-1\right)^2\,\Gamma \left(2-\frac{d}{2}\right) }{\Gamma (d-2)} \\
I^{12} & = & I^{21} = \frac{(3-d) } {k^2}  I^{11}\\ 
I^{22} & = & \frac{(3-d) (6-d) } {k^4}  I^{11}.
\end{eqnarray}
\end{subequations}
We remark that the UV divergences, which occurs only for even dimensions, arise from the factor 
$\Gamma \left(2-\frac{d}{2}\right)$ in $I^{11}$. 

Implementing the above described procedure as a straightforward computer algebra code, we readily 
obtain the following exact results for $C^{I}_1$ and $C^{I}_2$ 
\begin{subequations}
\begin{eqnarray}
  C^{\mbox{a}}_1 & = &
  \left[\frac{d-4}{8}  \xi^2 +\frac{3(d-4)}{4}\xi +\frac 1 2\frac{1}{d-1} -\frac{7d}{8}    \right]
  I^{11};\;\;
C^{\mbox{b}}_1  =   -\frac{1}{d-1} I^{11}
;\;\;C^{\mbox{a}}_2 = C^{\mbox{b}}_2  =  0
\end{eqnarray}
\end{subequations}
Adding the two contributions, we obtain the following {\it transverse} result for the one-loop contribution to $\langle BB \rangle$ 
\be\label{c9}
\Pi^{ab} _{\mu\nu}(p) = N g^2 \delta^{ab}
\left[\frac{d-4}{8}  \xi ^2 + \frac{3(d-4)}{4}  \xi + \frac{1}{2-2 d}-\frac{7 d}{8}\right] I^{11} 
\left(p_\mu p_\nu - p^2\eta_{\mu\nu}\right) .
\ee
We notice that this expression is gauge parameter independent only for $d=4$ in which case it has the following well known
UV pole for $d=4-2\epsilon$ \cite{abbott82}
\be\label{a16}
\left.\Pi_{\mu\nu}^{ab}(p)\right|_{UV}^{d=4} = -\frac{11}{3}\frac{N g^2}{16\pi^2\epsilon} i \delta^{ab} 
\left(p_\mu p_\nu - p^2\eta_{\mu\nu}\right),
\ee
which yields the correct gauge independent result for the beta function.

For $d=6-2\epsilon$, Eq. \eqref{c9} the UV pole becomes
\begin{eqnarray}\label{a17}
  \left.\Pi_{\mu\nu}^{ab}(p)\right|_{UV}^{d=6} &=&-
  \left(\frac{107}{240}-\frac{\xi}{8}-\frac{\xi^2}{48}\right) 
\frac{N g^2}{32\pi^3\epsilon}  i \delta^{ab} p^2  \left(p_\mu p_\nu - p^2\eta_{\mu\nu}\right) 
\nonumber \\
&=& - b_6  i \delta^{ab} p^2 \left(p_\mu p_\nu - p^2\eta_{\mu\nu}\right) ,
\end{eqnarray}
where 
\be\label{eqb6}
b_6 = \frac{N}{32\pi^3\epsilon}\left(\frac{107}{240}-\frac{\xi}{8} -\frac{\xi^2}{48}\right).
\ee

\begin{figure}
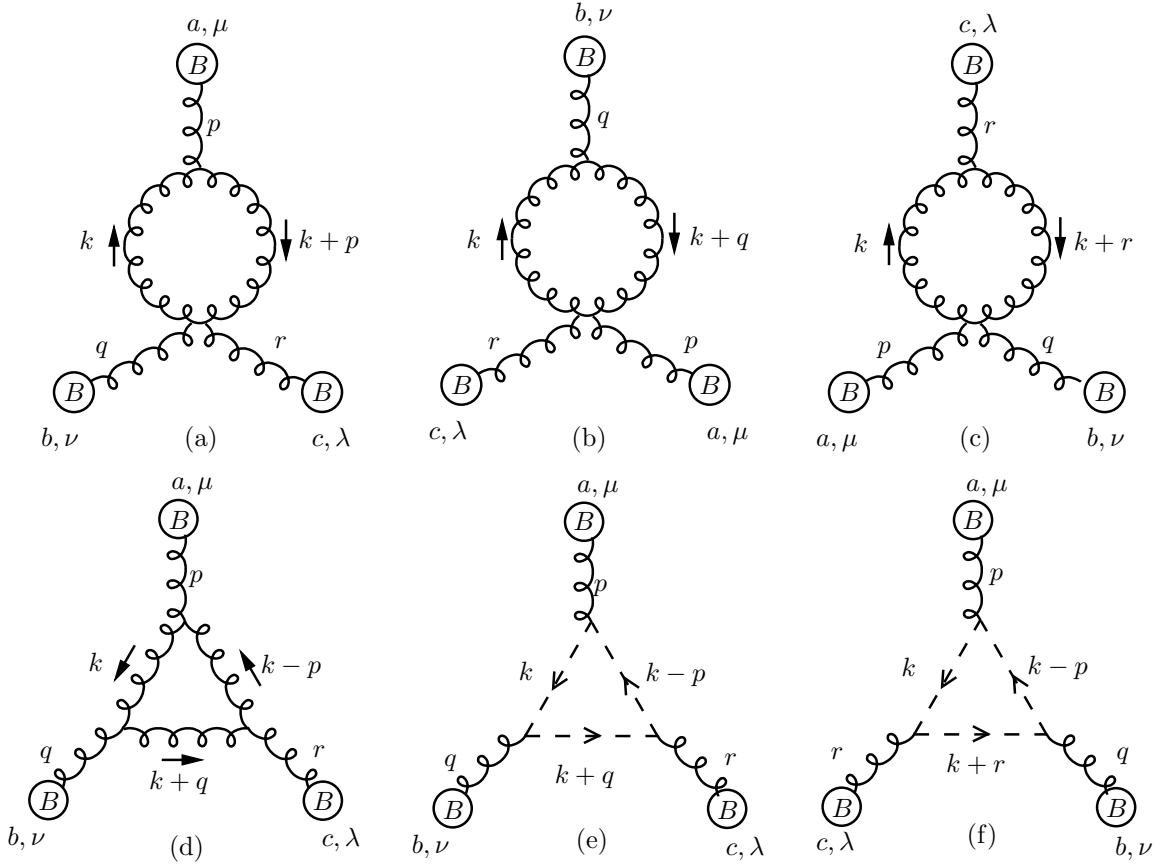

\begin{eqnarray}
\input{three4NewFix.pspdftex }  & \qquad \input{three6NewFix.pspdftex} & \qquad \input{three5NewFix.pspdftex} 
 \nonumber \\
\input{three1NewFix.pspdftex }  & \qquad \input{three2NewFix.pspdftex} & \qquad \input{three3NewFix.pspdftex} 
\nonumber
\end{eqnarray}
\caption{One-loop contributions to $\langle BBB \rangle$.}
\label{figThree}
\end{figure}

Let us now consider the three-gluon function. Figure \ref{figThree} shows the one-loop loop contributions
(one can easily verify that the contribution  obtained 
by joining the vertices \eqref{a5} and \eqref{a6} vanishes trivially). 

The calculation of graphs with two internal lines in 
figure \ref{figThree} (a), (b) and (c), is similar to the calculation of the self-energy graphs, in the sense that
the momentum integrals can be performed in a closed form, for any dimension $d$. This can be done using the
usual Feynman parametrization. We can also use the tensor decomposition 
procedure employed for the calculation of the self-energy, so that each tensor integral is reduced to the calculation
of scalars given by Eq. \eqref{scalInt}. Then, using the Feynman rules given by Eqs. \eqref{a1}, \eqref{a3} and \eqref{a4},
a straightforward computer algebra calculation yields the following
result for the diagram (a) in the figure \ref{figThree}
\be\label{a18}
{\Gamma{}^{(\mbox{{\scriptsize a}})} }^{abc}_{\mu\nu\lambda}(p,q,r) = 
\frac{\Gamma \left(\frac{d}{2}-1\right)^2\,\Gamma \left(2-\frac{d}{2}\right) }{\Gamma (d-2)} 
\frac{(\xi +3) (3 \xi +1) \left[(d-4)\xi  - d  \right]}{16 \xi }\frac{ N g^3 f^{abc} }{2^d\pi^{d/2}}
 (p^2)^{d/2-2} \left({p}_{\lambda } \eta_{\mu \nu }-{p}_{\nu } \eta_{\mu \lambda}\right).
\ee
The other two graphs can be obtained from cyclic permutations of \eqref{a18} as follows
\be\label{a19} 
{\Gamma{}^{(\mbox{{\scriptsize b}})}}^{abc}_{\mu\nu\lambda}(p,q,r) =
{\Gamma{}^{(\mbox{{\scriptsize a}})}}^{abc}_{\nu\lambda\mu}(q,r,p)   \;\;\;
\mbox{and} \;\;\;
{\Gamma{}^{(\mbox{{\scriptsize c}})}}^{abc}_{\mu\nu\lambda}(p,q,r) =
{\Gamma{}^{(\mbox{{\scriptsize a}})}}^{abc}_{\lambda\mu\nu}(r,p,q)   .
\ee
The expression \eqref{a18} has UV divergences for even dimensions. In particular, for $d=4$ and $d=6$ we obtain 
\be\label{a20}
\left.{\Gamma{}^{(\mbox{{\scriptsize a}})} }^{abc}_{\mu\nu\lambda}(p,q,r)\right|_{UV}^{d=4} = 
-\frac{(\xi +3) (3 \xi +1)}{4 \xi }\frac{ N g^3 f^{abc} }{16 \pi^2\epsilon} 
\left({p}_{\lambda } \eta_{\mu \nu }-{p}_{\nu } \eta_{\mu \lambda}\right)
\ee
and
\be\label{a21}
\left.{\Gamma{}^{(\mbox{{\scriptsize a}})} }^{abc}_{\mu\nu\lambda}(p,q,r)\right|_{UV}^{d=6} = 
-\frac{(\xi^2 -9) (3 \xi +1)}{96 \xi }
\frac{ N g^3 f^{abc} }{32 \pi^3\epsilon} (p)^2
\left({p}_{\lambda } \eta_{\mu \nu }-{p}_{\nu } \eta_{\mu \lambda}\right). 
\ee
In four dimensions, the full result (adding the two permutations in figures \eqref{figThree} (b) and (c), becomes proportional to the tree three
gluon vertex

\be\label{a20tot}
\left.{\Gamma{}^{(\mbox{{\scriptsize a+b+c}})} }^{abc}_{\mu\nu\lambda}(p,q,r)\right|_{UV}^{d=4} =
-\frac{(\xi +3) (3 \xi +1)}{4 \xi }\frac{ N g^2 }{16 \pi^2\epsilon} 
g f^{abc} \left[\eta_{\mu\lambda}({r}_\nu-{p}_\nu) 
                                                                         +\eta_{\mu\nu}({p}_\lambda-{q}_\lambda) 
                                                                   +\eta_{\nu\lambda}({q}_\mu-{r}_\mu)\right]  .
\ee

Let us now consider  graphs (d), (e) and (f) of Fig. \ref{figThree}. To illustrate the method of calculation, 
we first consider the ghost loop diagram in  Fig \ref{figThree} (e). Using
Eqs. \eqref{a2} and \eqref{a5}, we can express this contribution as follows
\begin{eqnarray}\label{a22}
-{i N g^3 f^{abc}}\left\{
4 J^{\mu\nu\lambda}_{111} + 2\left[J^{\mu\nu}_{111}({q}_\lambda-{p}_\lambda) 
+ J^{\mu\lambda}_{111} {q}_\nu  -  J^{\nu\lambda}_{111} {p}_\mu \right]
\right. \nonumber \\ \left.
- J^\lambda_{111} {p}_\mu {q}_\nu 
- J^\mu_{111} {q}_\nu( {p}_\lambda -  {q}_\lambda)
+ J^\nu_{111}  {p}_\mu( {p}_\lambda -  {q}_\lambda)
+ J_{111} \frac{{p}_\mu {q}_\nu( {p}_\lambda -  {q}_\lambda)}{2}
\right\},
\end{eqnarray} 
where
\be\label{a23}
J^{\mu_1\mu_2\cdots\mu_j}_{l m n } \equiv 
\int \frac{d^d k} {(2\pi)^d}
\frac{k^{\mu_1 } k^{\mu_2}\cdots k^{\mu_j } }{(k^2+ i 0)^l [(k-p)^2+i 0]^m [(k+q)^2+i 0]^n}
\ee
is the general type of integrals that will also arise in the much more involved contribution from the
diagram of Fig. \ref{figThree} (d) (terms which depend on $\xi$ will involve integrals with $m$, $n$ and $l$ greater than one).

Since the integrals in Eq. \eqref{a23} cannot be expressed in a closed form as a function of the dimension $d$
(the integrals in the two Feynman parameters cannot be done in a closed form, for a general $d$), we must
now first expand in $\epsilon=(n-d)/2$ and afterwards perform the Feynman parameter integrals. In particular, when
considering the UV pole $1/\epsilon$, the Feynman parameter integrals become trivial.

For a given dimension $n$ such that $d=n-2 \epsilon$ not all the integrals in Eq. \eqref{a23} will be divergent. For instance,
for $n=4$ only the terms in the first line of Eq. \eqref{a22} are UV divergent. By power counting, these divergences come from 
$J^{\mu\nu\lambda}_{111}$, $J^{\mu\lambda}_{111}$  and $J^{\nu\lambda}_{111}$.  Computing these integrals with the standard Feynman 
parametrization procedure, and adding also the contribution from the graph in the Fig. \ref{figThree} (f),
we obtain the following pole part for the ghost loop diagrams in four dimensions
\be\label{a24}
\left.{\Gamma{}^{(\mbox{{\scriptsize e+f}})} }^{abc}_{\mu\nu\lambda}(p,q,r)\right|_{UV}^{d=4} =
-\frac 1 3 \frac{g^2 N }{16 \pi^2 \epsilon} g f^{abc} \left[\eta_{\mu\lambda}({r}_\nu-{p}_\nu) 
                                                                         +\eta_{\mu\nu}({p}_\lambda-{q}_\lambda) 
                                                                   +\eta_{\nu\lambda}({q}_\mu-{r}_\mu)\right]  ,
\ee
which is proportional to the tree three gluon vertex. Using the same basic procedure, we have obtained the following result
for the graph in Fig. \ref{figThree} (d) in four dimensions
\be\label{a24a}
\left.{\Gamma{}^{(\mbox{{\scriptsize d}})} }^{abc}_{\mu\nu\lambda}(p,q,r)\right|_{UV}^{d=4} =
-\frac{9 \xi ^2-10 \xi +9}{12 \xi }
\frac{g^2 N }{16 \pi^2 \epsilon} g f^{abc} \left[\eta_{\mu\lambda}({r}_\nu-{p}_\nu) 
                                                                         +\eta_{\mu\nu}({p}_\lambda-{q}_\lambda) 
                                                                   +\eta_{\nu\lambda}({q}_\mu-{r}_\mu)\right]  
\ee
Adding Eqs. \eqref{a20tot}, \eqref{a24} and \eqref{a24a}, the gauge parameter dependence cancells and we are left with the following result in four dimensions
\be\label{a24b}   
\left.{\Gamma }^{abc}_{\mu\nu\lambda}(p,q,r)\right|_{UV}^{d=4} =
-\frac{11}{3}
\frac{g^2 N }{16 \pi^2 \epsilon} g f^{abc} \left[\eta_{\mu\lambda}({r}_\nu-{p}_\nu) 
                                                                         +\eta_{\mu\nu}({p}_\lambda-{q}_\lambda) 
                                                                   +\eta_{\nu\lambda}({q}_\mu-{r}_\mu)\right]  
\ee

In six dimensions, all the terms in \eqref{a22} have a UV pole. As a consequence, the tensor structure of the resulting expression is
much more involved than in four dimensions. In the case of the graph in Fig. \ref{figThree} (d), because of the gauge parameter dependence,
the possible structures in the integrand are even richer, so that there will be terms
involving integrals like
$J_{111}$, $J^\mu_{111}$, $J^{\mu\nu}_{112}$, $\dots$, $J_{112}^{\mu_1 \mu_2 \dots \mu_7}$, \dots, $J_{122}^{\mu_1 \mu_2 \dots \mu_7}$ (the order of
possible UV divergences goes up to three). We have generated all the needed integrals using a computer algebra code. After pattern matching all the possible terms
in the integrand and making the corresponding substitutions, we have obtained the following result in six dimensions
\begin{eqnarray}\label{a25}
  \left.{\Gamma }^{abc}_{\mu\nu\lambda}(p,q,r)\right|_{UV}^{d=6} &=&\frac{Ng^3}{16 \pi^3\epsilon} f^{abc}\left\{
  \left(-\frac{\xi ^2}{48}-\frac{\xi }{8}+\frac{107}{240}\right)\left[
    {p}^{2} (-{p}_{\nu } g_{\lambda \mu }- {r}_{\mu } g_{\lambda \nu }+ {r}_{\nu } g_{\lambda \mu })
\right.\right. \nonumber \\ && \left.\left.
+ {p}\cdot {r} \,{p}_{\mu } g_{\lambda \nu } 
+{p}_{\lambda }{p}_{\mu } ({p}_{\nu }- {r}_{\nu })\right]
  +
  \right. \nonumber \\ 
&& \left. \frac{1}{180}\left[
p\cdot q \left({r}_{\mu } \eta_{\lambda \nu }-{r}_{\nu } \eta_{\lambda \mu }\right)+
p\cdot r \left({q}_{\lambda } \eta_{\mu \nu }-{q}_{\mu } \eta_{\lambda \nu }\right)+ \right.\right. \nonumber \\ 
&& \left.\left.  q\cdot r \left({p}_{\nu } \eta_{\lambda \mu }-{p}_{\lambda } \eta_{\mu \nu }\right)+
{p}_{\lambda } {q}_{\mu } {r}_{\nu }-{p}_{\nu } {q}_{\lambda } {r}_{\mu }\right]
  \right. \bigg\}
\nonumber \\ &&
  + \;\mbox{six permutations of}\; (a,\mu,p),\;(b,\nu,q),\;(c,\lambda,r) .
\end{eqnarray}

We now consider Eqs. \eqref{a16}, \eqref{a17}, \eqref{a24b} and \eqref{a25} as the effective ``vertices''  which can be read from effective actions of the form
\be\label{a26a} 
S_4  = a_4 \int d^4 x (F^{a}_{\mu\nu})^2 
\ee
and
\be\label{a26}
S_6 =  \int d^6 x \left[ b_6 [(D F)_\mu^a]^2 + a_6  f^{abc} F^a_{\mu\nu} F^{b\nu}_\rho F^{c\rho\mu} \right] ,
\ee
where 
\be\label{a27}
F^{a}_{\mu\nu} = \partial_\mu B^a_\nu - \partial_\nu B^a_\mu  + g f^{abc}  B^b_\mu B^c_\nu
\ee
and
\be\label{a28}
(DF)_\nu^a =  \partial^\mu F_{\mu\nu}^a + g f^{abc} B^{b\mu} F_{\mu\nu}^c.
\ee
Comparing the quadratic and cubic parts of  $S_4$ and $S_6$, in momentum space, with  Eqs. \eqref{a16}, \eqref{a17}, \eqref{a24b} and \eqref{a25} we obtain
\be\label{a29}
a_4 = - \frac{11}{3} \frac{N}{16 \pi^2\epsilon},
\ee
where $b_6$ is the same coefficient defined in Eq. \eqref{eqb6} and
\be\label{a30}
a_6 =\frac{N}{32\pi^3\epsilon}\frac{1}{180}
\ee
is a gauge parameter independent coefficient.
It is remarkable that all the gauge parameter dependence is only present in the first term of $S_6$, which would vanish on-shell.


\section{}

\begin{figure}[t]
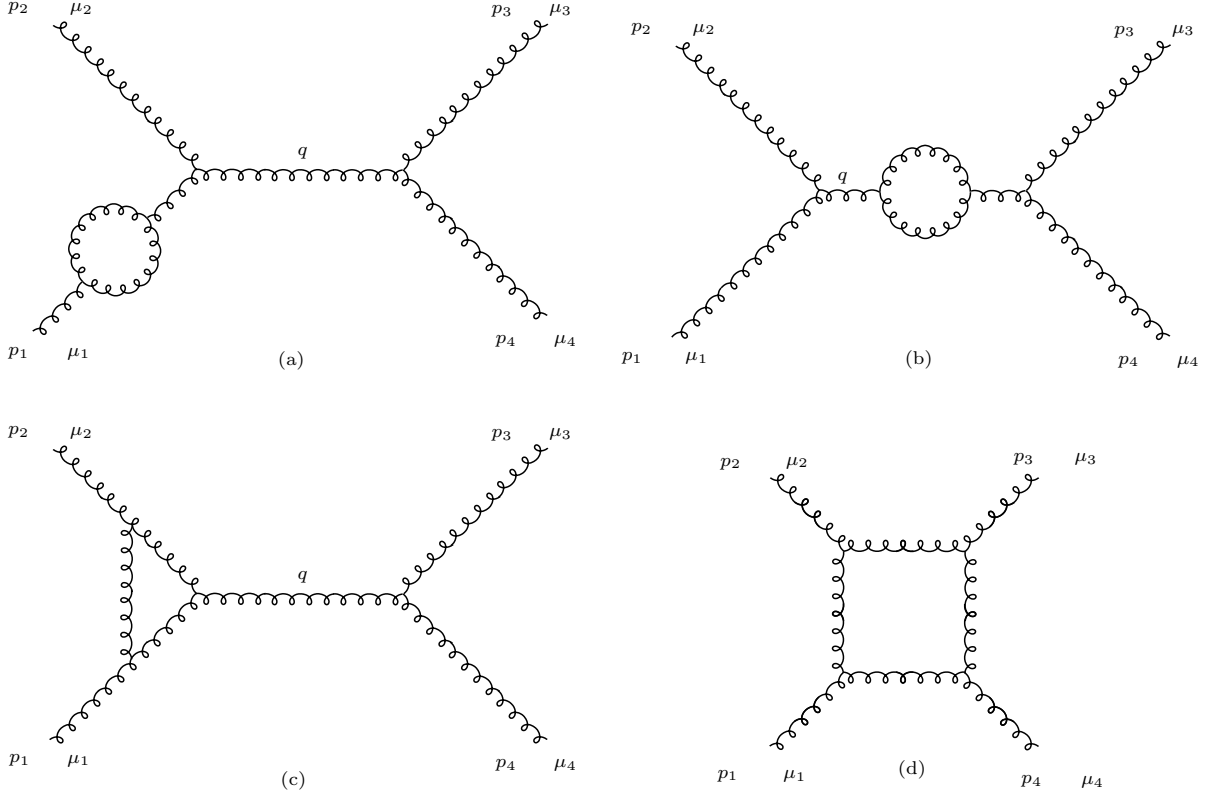

\begin{eqnarray}
  \input{ampa.pspdftex}   & \qquad   \input{ampb.pspdftex} 
 \nonumber \\ & \nonumber \\
\input{ampc.pspdftex }   & \qquad \input{ampd.pspdftex}
\nonumber
\end{eqnarray}
  \caption{Examples of one loop Feynman diagrams for the gluon-gluon scattering amplitude.}\label{fig2}
\end{figure}

We show here that the gauge-independence of the coefficient $c_{12}$ in 
\eqref{eq10} ensures the gauge invariance of the one-loop S-matrix elements. 
To this end, let us consider the gluon-gluon scattering 
amplitude $p_1 + p_2 \rightarrow p_3 + p_4$ involving a pair of background fields which are on-shell,  which 
contains transverse polarization vectors. To one-loop order, typical Feynman diagrams 
are shown in Fig. \ref{fig2}, but other relevant graphs must also be included.

The contribution from the graph Fig \ref{fig2} (a) involves a background gluon self-energy which has a 
form consistent with that coming from the structure proportional to $c_{11}$ in \eqref{eq9}. 
This yields a factor 
\be\label{b1}
c_{11} g^2 (p_1^2\eta^{\mu_1 \sigma} - p_1^{\mu_1} p_1^\sigma)
\ee
which vanishes on shell, at $p_1^2 = 0$, when contracted with a transverse polarization vector.
However, the self-energy insertion on the internal line shown in graph \ref{fig2} (b) is non-vanishing.
But 
it may be verified that the contributions arising from the structure $(D_\sigma F^{\sigma\mu})^2$ 
cancel out among themselves in the sum of the graphs
in Fig \ref{fig2} (b),(c) and (d). Thus, the first counterterm in Eq. \eqref{eq9}  has no observables effects, as expected. 
On the other hand, the structure $F_\mu^{\;\;\sigma}\cdot (F_{\sigma\rho}\wedge F^{\rho\mu})$
in the second counterterm in \eqref{eq9}, 
does contribute  to the diagrams shown in figures in \eqref{fig2} (c) and (d). For example, diagram (c) will yield
in this case a contribution involving the factor
\be
c_{12} g^2 (p_1^\sigma \eta^{\rho\mu_1} - p_1^\rho \eta^{\sigma\mu_1})(p_2^\rho \eta^{\tau\mu_2} - p_2^\tau \eta^{\rho\mu_2})(q^\tau \eta^{\sigma\alpha} - q^\sigma \eta^{\tau\alpha})
\ee
which is transverse with respect to the momenta $p_1^{\mu_1}$, $p_2^{\mu_2}$, $q^{\alpha}$. 
This is a reflection of the Ward identity satisfied by the last structure in Eq. \eqref{eq9}.
Then, the   $\xi$    dependent part of the gluon propagator 
\be
\frac{1}{q^2}\left[\eta_{\alpha\beta} - (1-\xi) \frac{q_\alpha q_\beta}{q^2}\right]
\ee
decouples. Thus, a possible gauge dependence may come only from the coefficient $c_{12}$. 
A similar conclusion also holds for the corresponding contribution arising from diagram (d).
But the sum of such contributions yields the S matrix element for the gluon-gluon 
scattering amplitude at one loop order, which should be gauge-invariant. This requires 
the coefficient $c_{12}$  to  be a gauge-independent quantity.

\section{}

We discuss here some useful Bianchi identities for the YM fields, which hold in any dimension,
whether or not the gauge fields satisfy the field equations. We start from the identity 
\be\label{c1}
D_\rho { F}_{\mu\nu}+D_\mu { F}_{\nu\rho}+D_\nu { F}_{\rho\mu} = 0.
\ee
Multiplying \eqref{c1}  by   ${ F}_{\mu\nu} D_\rho$  and using the anti-symmetry of ${ F}_{\mu\nu}$   under $\mu\leftrightarrow\nu$, we obtain
\be\label{c3}
         { F}_{\mu\nu} D^2 { F}^{\mu\nu} 
         + 2 { F}^{\mu\nu} \cdot D_\rho D^\mu { F}^{\nu}_{\;\rho} = 0. 
\ee
Employing the relation                                     
\be\label{c4}
D_\mu D_\nu - D_\nu D_\mu = - g \, { F}_{\mu\nu}
\ee
one can write \eqref{c3} in the form
\be\label{c5}
         { F}_{\mu\nu} D^2 { F}^{\mu\nu} 
         + 2 { F}_{\mu\nu} \cdot D^\mu D_\rho { F}^{\nu\rho} +  g   { F}_{\mu\nu} \cdot { F}^{\mu\rho} \wedge { F}_{\rho}^{\;\;\nu} = 0. 
\ee
Integrating by parts the second term in  \eqref{c5}, we get the identity
\be\label{c6}
         (D^\mu { F}_{\mu\nu})\cdot (D_\rho { F}^{\rho\nu}) + \frac 1 2 { F}_{\mu\nu} D^2 { F}^{\mu\nu}  
+  g { F}_{\mu\nu} \cdot { F}^{\mu\rho} \wedge{ F}_{\rho}^{\;\;\nu} = \partial^\mu({ F}_{\mu}^{\;\;\nu} \cdot {D}^{\rho} { F}_{\nu\rho}).
\ee
Since a total derivative may be disregarded, we see that only two of the above three structures can be taken as being independent.
These two structures may then be used for the gauge-invariant counter-terms required to one-loop order.

In order to find the relevant  structures at two-loops, we multiply \eqref{c1}, for example, by 
\be\label{c7}
         { F}_{\mu\sigma}  { F}^{\sigma\nu} D^\rho;\;\;
         { F}_{\mu\nu}  { F}^{\rho\sigma} D_\sigma;\;\;
         { F}_{\mu\sigma} D^\rho  { F}^{\sigma\nu};\;\;
         { F}_{\mu\nu} D_\sigma  { F}^{\rho\sigma};\;\;
         { F}_{\mu\nu} D_\sigma D_\rho D^\sigma;\;\;
         { F}_{\mu\sigma} D_\nu  D^\sigma D^\rho;\;\;
         \ee
Ignoring total derivatives and using a procedure similar to the one employed above, it turns out that the Bianchi identities lead to the following independent structures
\begin{eqnarray}\label{c8}
&\left[D_\mu D_\sigma{ F}^{\sigma\nu}\right]^2;\;\;
(D_\sigma{ F}^{\sigma\mu})\wedge (D_\rho{ F}^{\rho\nu})\cdot { F}_{\mu\nu};\;\;
(D^\mu D^\sigma{ F}_{\sigma\nu}) \cdot ( { F}^{\nu\rho} \wedge { F}_{\rho\mu});\;\;
\nonumber \\
& [({ F}_{\mu\nu})^2]^2;\;\;
\left[{ F}_{\mu\sigma}  \cdot { F}^{\sigma\nu}\right]^2;\;\;
\left[{ F}_{\mu\nu}  { F}^{\rho\sigma}\right]^2;\;\;
     ({ F}_{\mu\sigma} \wedge { F}^{\sigma\rho})\cdot ({ F}_{\rho\nu} \wedge { F}^{\nu\mu}).
\end{eqnarray}
We observe that the first three structures vanish when the equations of motion are used.
Thus, on shell, four new gauge-invariant counter-terms are required at the two-loop order.

\section{}

We examine here the dependence of the beta function on the definition of the coupling constant.
To this end, consider two definitions $\tilde g(\mu)$ and $\bar g(\mu)$
of the running coupling corresponding to different definitions of the renormalization scale $\mu$.
Since both couplings  are finite and dimensionless, we must have      $\bar g(\mu) = \bar g[\tilde g(\mu)] $,
so that
\be\label{d1}
\bar\beta(\bar g) \equiv \mu \frac{d\bar g}{d\mu} = \mu \frac{d\bar g}{d\tilde g} \beta(\tilde g)
\ee
Using the same definition of the bare coupling constant, the  renormalized couplings must be 
equal to lowest order, so one may write
\be\label{d2}
\bar g(\tilde g) = \tilde g + b {\tilde g}^3 + {\cal O}({\tilde g}^5)
\ee
or, equivalently
\be\label{d3}
\tilde g = \bar g -  b {\bar g}^3 + {\cal O}({\bar g}^5)
\ee
In $d$ dimensions, the power series for $\beta(\tilde g)$ takes the form $[c=(d-4)/2]$
\be\label{d4}
\beta(\tilde g) = c \tilde g + c^\prime {\tilde g}^3 + {\cal O}({\tilde g}^5)
\ee
This may be re-written in terms of   $\bar g$, as  
\be\label{d5}
\beta(\bar g) = c \bar g + (c^\prime-b c)  {\bar g}^3 + {\cal O}({\bar g}^5)
\ee
From \eqref{d1} and \eqref{d3} we have then 
\begin{eqnarray}\label{d6}
\bar\beta(\bar g) &=& \left[1 + 3 b {\bar g}^2 + {\cal O}({\bar g}^4)  \right]\left[c {\bar g} + (c^\prime - bc){\bar g}^3 + {\cal O}({\bar g}^5)  \right]
\nonumber \\
&=& c {\bar g} + (c^\prime + 2 bc){\bar g}^3 + {\cal O}({\bar g}^5).  
\end{eqnarray}
We see that the first term in the power series for $\bar\beta$
in terms of  $\bar g$     has the same coefficient as in the power series for $\beta$  in terms of
$\tilde g$. But this is generally  not the case for higher order
terms in the power series of the beta function.
For instance, one could choose in \eqref{d2} the coefficients of higher powers of
$\tilde g$, such that all higher order terms  in  \eqref{d6}         would vanish.

\begin{acknowledgments}
F. T. B. and J. F. would like to thank CNPq for financial support. D. G. C. M. would like to thank Roger
Macleod for an enlightening discussion, the Universidade de S\~ao Paulo for the hospitality
during the realization of this work and FAPESP for financial support (grant number 2018/01073-5).
\end{acknowledgments}


\end{document}